\documentclass[a4paper,12pt]{article}
\usepackage{epsfig,amsmath,amssymb,pifont,verbatim,natbib}
\usepackage{setspace}
\newcommand{\ud}{\mathrm{d}}
\newcounter{theorem}
\newtheorem{definition}[theorem]{Definition}
\title{Statistical Significance Revisited}
\author{Reason L. Machete}
\author{Machete, R. L.\\
{\footnotesize Botswana International University of Science and Technology}}
\date{May 4, 2026}
\begin{document}
\maketitle
\noindent\rule{14.9cm}{0.4pt}
\begin{abstract}
Since its introduction by Fisher, the method of hypothesis testing that relies on computing error probabilities has witnessed several developments. Perhaps the most significant development was the seminal contributions of Neyman and Pearson who brought in the concept of the alternative hypothesis with its corresponding error of the second kind. Significance tests have played a major role in various scientific and technological developments, but not without controversies. Although originally cast as frequentist approaches, Bayesian ideas have been incorporated into significance tests, widening access to them. The quantities central to computations of error probabilities are the sampling distributions, which can be computed even without thresholds or alternative hypotheses. Even though Fisher used the significance threshold of 0.05 in his calculations, he cautioned against prescribing any specific threshold~\citep{fisher-1955,fisher-1971}. Recently, there have been calls for reformation in practice with regard to the almost standard use of the significance threshold of 0.05, prepublication confirmatory studies, the dichotomous consideration of the null and alternative hypothesis and abandoning significance tests altogether in favour of other approaches such as confidence intervals and Bayesian decision theory. In this paper, we examine these calls for reform and unearth their strengths and short comings.
\end{abstract}
\noindent\rule{14.9cm}{0.4pt}
\section{Introduction}
Significance tests can be traced to the works of Fisher~\citep{fisher-1925,fisher-1936,fisher-1955,fisher-1971}, Neyman~\citep{neyman-1955} and Pearson~\citep{neyman-1928,neyman-1928b,neyman-1933}. Wald also made significant contributions to the broader field of decision analysis within a short space of three years, from 1937 to his demise in 1940~\citep{neyman-1955}. Fisher may be considered the pioneer of this field, and his original work dealt with scenarios where there was only one hypothesis, the {\it null hypothesis}~\citep{fisher-1925}. He developed significance tests for the null hypothesis, the  aim being to control deviations from it when it is correct: This is an error probability, and it is called a type I error. An important threshold for this is the significance level, $\alpha$, and it is used to control the type I error. Computed probabilities, $p$-values, are compared with the threshold to make inferences in scientific investigations. 

Neyman and Pearson built on the earlier work of Fisher, introducing an alternative hypothesis and arguing that another error needs to be controlled~\citep{neyman-1933,neyman-1955}: that is the error of finding no deviations from the hypothesis when it is incorrect. The introduction of the alternative hypothesis dismayed Fisher, who accused Neyman and Pearson of being in a mental state of confusion~\citep{fisher-1955}. Their first works on this appeared in 1928~\citep{neyman-1928,neyman-1928b}, but their collaboration is thought to have started in 1926~\citep{mayo-2018}. Without the alternative hypothesis, it would not be possible to compute the probability of the type II error (or error of the second kind, as they called it). Along side this, they also introduced the power of the test, a measure of how well the test is able to detect this error. A method is powerful if it minimises the probability of a type II error, $\beta$. In this case $(1-\beta)$ is the power of the test. The alternative hypothesis is often a compliment of the null hypothesis. Whereas Fisher spoke of significance, Neyman and Pearson introduced the term acceptance~\citep{neyman-1933}. Mayo calls one who is concerned with controlling error probabilities an error statistician~\citep{mayo-2018}. Among the error statisticians, there is the severe tester to whom induction is the procedure for corroborating by severe testing, a concept discussed in detail in~\cite{mayo-2018}. This is different from the Fisherian definition of induction, who thinks of inductive logic (inference) in the sense of reasoning from the ``sample to the population from which the sample came from, from the consequences to the causes, or from the particular to the general"~\citep[][page 69]{fisher-1955}. 

Meanwhile, to make inferences, the Bayesian and Likelihoodist tend to focus on the Bayes factors and likelihood ratios, which are comparative measures of evidence. Given a hypothesis, $H$ and data $\boldsymbol{x}$, the Bayesian is primarily concerned with updating his prior belief $\mbox{Pr}(H)$ using the likelihood to obtain the posterior probability $\mbox{Pr}(H|\boldsymbol{x})$, that is:
\begin{equation}
\mbox{Pr}(H|\boldsymbol{x})=\frac{\mbox{Pr}(\boldsymbol{x}|H)}{\mbox{Pr}(\boldsymbol{x})}\mbox{Pr}(H).
\end{equation}
As Sprott put it~\citep[][page 331]{lindley-2000},``Bayes's theorem requires all possible hypotheses to be specified in advance, along with their prior probabilities. Any new, hitherto unthought of hypothesis or concept will necessarily have a prior probability of zero." Bayesians often argue that the $p$-value exaggerates evidence since it often yields values that are lower than the posterior probability~\citep{mayo-2018}. Inspite of this, the $p$-value remains widely used to make inferential statements.
 
Problems of reproducibility and replication in science and misinterpretations of $p$-values culminated in the American Statistical Association (ASA) issuing a statement of $p$-values~\citep{wasserstein-2016}. A year later,~\cite{benjamin-2017} called for a change in practice from using a significance threshold of 0.05 to use 0.005 instead, to bolster replication rates, at least for reports of novel findings. They suggested that novel results should otherwise be reported as merely indicative when the threshold used is 0.05. It seemed that such a change had potential to double the replication rates, at least in the specific disciplines considered. As an antidote to the replication problem, ~\cite{nosek-2012} suggested a two part approach whereby the first study is exploratory and the second study confirmatory. This approach was later emphasised by~\cite{gelman-2014} as one way to address noted problems of multiple comparisons. 

In 2019, the ASA's statement on $p$-values was followed up by a series of papers under the theme, Statistical Inference in the 21st Century~\citep{wasserstein-2019}. The authors were called to provide different perspectives of a {\it A world beyond $P<0.05$}.~\cite{colquhoun-2019} urged that $p$-values and confidence intervals should continue to be provided and be supplemented with the {\it false positive risk}. In order to compute the false-positive risk, he suggested using a prior of the null hypothesis of $0.5$.~\cite{gannon-2019} called for retaining the useful concept of statistical significance, and to use tests that are compatible with the likelihood principle. Meanwhile the likelihood principle has the disdvantage of being indifferent to optional stopping.~\cite{ioannidis-2019} dissuades the use of $p$-values in the majority of studies, offering effect sizes, confidence intervals, methods based on false discovery rates, Bayesian methods, and more stringent thresholds as alternatives to help stem the highlighted tide of selection bias.~\cite{hubbard-2019} proposed contemplating abandoning the use of $p$-values in null hypothesis significance testing.~\cite{mcshane-2019} proposed abandoning statistical significance and reporting the $p$-value continuously without thresholds. Instead of statistical significance, they suggested using {\it currently} subordinate factors, which include prior evidence, plausibility of mechanism, and real world costs and benefits. They deem their proposal to be consistent with Principle 3 of the ASA statement that states that "Scientific conclusions and business or policy decisions should not be based only on whether a p-value passes a specific threshold." Reporting $p$-values in a continuous way is re-iterated by~\cite{amrhein-2019b}, who also suggest augmenting it with compatibility intervals. In tune with this is the sentiment of ~\cite{amrhein-2019} to retire statistical significance altogether and use confidence intervals.~\cite{greenland-2019} called for avoiding dichotomising the $p$-value or comparisons of $p$-value to $\alpha$ levels, among others, citing the works of~\cite{poole-1987}, ~\cite{hoekstra-2006} and~\cite{wasserstein-2016} in his favour.~\cite{benjamin-2019} recommended that the significance threshold should be changed from 0.05 to 0.005, reiterating the call they made earlier~\citep{benjamin-2017}. The cornerstone of this call is the diagnostic screening formula first suggested by~\cite{browner-1987}, the false positive risk recommended by~\cite{colquhoun-2019} and one minus the positive predictive value popularised by~\cite{ioannidis-2005}.

In summary, the calls for reform are as follows: i) reduce the significance threshold ii) prepublication of confirmatory studies iii) abandon significance with its associated thresholds iv) report $p$-values in a continuous way v) use confidence intervals and vi) use costs and benefits analysis. This paper examines these calls for reform, evaluates the supporting evidence and highlights some implications of heeding them. The merits and shortcomings of reducing the threshold are addressed in light of the false discovery rate (or equivalently, positive predictve value) and cost function analysis. Implications of $p$-value distributions in assessing the strength of evidence and reproducibility of results are discussed. The call to retire statistical significance and avoid dichotomisation hits at the very foundation of Fisherian and Neyman-Pearson tests and leaves the decision making to be done only with confidence intervals and some ad-hoc approaches. Please note that Fisher's work on significance tests contained fiducial intervals, albeit with the blemish of probabilistic instanciation~\citep{fisher-1955}. The next section presents the basics of Neyman and Pearson tests and the effects of changing the significance threshold on replications rates. Section~\ref{sec:basis} presents the basis for the call to reduce the significance threshold. In Section~\ref{sec:analysis}, an analysis of the effect of threshold reduction on replication rates is given. Alternative reforms are discussed in Section~\ref{sec:others}, including the severity construals advocated for by~\cite{mayo-2018}. Section~\ref{sec:pvalues} discusses $p$-value distributions and their implications to reproducibility and strength of evidence. In Section~\ref{sec:examples}, sub-Saharan Africa savanna fires~\citep{machete-2023} are used as an illustrative example of the use of significance tests. The paper concludes with a discussion of the key points in Section~\ref{sec:discussion}.
\section{Error Probabilities}
 In order to introduce the N-P error statistics, let us suppose that $\boldsymbol{X}$ is a random variable, of which the data $\boldsymbol{x}$ is a particular realisation. Associated with the test will be the test statistic, $\ud(\boldsymbol{X})$. The statistic should be such that the larger it is, the further the outcome is from what is expected under the null hypothesis $H_0$ with respect to the question at hand. In this case, the $p$-value can be computed as $p(\boldsymbol{x})=\mbox{Pr}(\ud(\boldsymbol{X})\ge\ud({\boldsymbol{x}});H_0)$. It is the probability that the test would have produced a result at least as large as the one observed, under the null hypothesis $H_0$~\citep[see][]{mayo-2022}. Borrowing from Fisher~\citep{fisher-1955}, we can think of the parameter $\lambda$ as being governed by a parametric model $\mbox{M}_\lambda(\boldsymbol{x})$ defined by
\begin{equation}
\mbox{M}_\lambda(\boldsymbol{x})=\left\{f(\boldsymbol{x};\lambda)|\lambda\in\Lambda\right\}.
\end{equation}
where $f(\boldsymbol{x};\lambda)$ is the sample distribution and $\Lambda$ denotes the set of all possible values of the unknown parameter, $\lambda$. The {\it null} and {\it alternative} can then be represented as
\begin{equation}
H_0:\lambda\in\Lambda_0\quad\mbox{vs.}\quad H_1:\lambda\in\Lambda_1,
\end{equation}
where $\Lambda_0\subset\Lambda$, $\Lambda_1\subset\Lambda$, $\Lambda_0\cap\Lambda_1=\emptyset$ and $\Lambda_0\cup\Lambda_1=\Lambda$. Neyman and Pearson called $H_0$ the {\it test hypothesis}. The type I error probability is then fixed according to the relation
$$\mbox{Pr}(\ud(\boldsymbol{X})\ge c_\alpha;H_0)\le\alpha,$$
where $\alpha$ is the significance level, often taken to be 0.05. Whenever the $p$-value is below this threshold, the available data are taken as evidence against $H_0$. In this sense, the $p$-value can be thought of as the probability of erroneous rejection of $H_0$, or the {\it false positive}. The null is rejected whenever the $p$-value is less than $\alpha$. On the other hand, the type II error probability is given by
$$\mbox{Pr}(\ud(\boldsymbol{X})<c_\alpha;\theta)=\beta(\theta),$$
where $\theta$ is in the region of the alternative hypothesis $H_1$. This is the probability of failing to reject the null hypothesis when it is false. It is related to the {\it power} of the test, which we can denote by $1-\beta$. The most powerful test is the one that minimises $\beta$, thereby maximising $1-\beta$.
\subsection{The Basis of Threshold Reduction}
\label{sec:basis}
Recently there have been calls for reform with regard to significance tests. Here, the basis of the call to use a more stringent significance threshold is examined.~\cite{benjamin-2017} called for the significance level to be adjusted to $\alpha = 0.005$ instead of the standard $\alpha=0.05$, at least for claims of new discoveries.~\cite{ioannidis-2019} also supported this call. Their main reason for this call for reform is to improve the replication rate of scientific studies. In Psychology and in Experimental Economics, they found replication rates for studies with $p<0.005$ to be double those where $0.005<p<0.05$. A Bayesian argument that the ratio of the posterior probability of $H_1$ against that of $H_0$ is a more direct measure of the strength of evidence\footnote{Seemingly also implying more relevant} is used to support the use of the Bayes Factors over $p$-values. Admitting that there is no direct link between Baye's Factors and $p$-values since the Bayes factor depends on $H_1$, the authors go on to argue that the $p$-value exaggerates the evidence.  This is an old argument that is confounded with the idea that posterior probabilities are superior to error probabilities. Their arguments were  further supported by hypothetical calculations of the false positive rate, using the formula
\begin{equation}
\mbox{False positive rate} \approx \frac{\alpha\varphi}{\alpha\varphi+(1-\beta)(1-\varphi)},
\label{eqn:fpr}
\end{equation}
where $\varphi$ is the probability that the null hypothesis is true, i. e. $\varphi= \mbox{Pr}(H_0)$,  which~\cite{benjamin-2017} approximate by the proportion of null hypotheses that are true. The supplementary information to their paper gives details of how the formula was arrived at. Their derivation appeals to the strong law of large numbers and the continuous mapping theorem. This is essentially the same formula proposed by~\cite{ioannidis-2005} for diagnostic screening. It is a Bayesian approach that can be traced back to~\cite{browner-1987}, who argued that the $p$-value, power and confidence intervals are necessary for interpreting the results of a research study, but not sufficient: the scientist must estimate the prior probability that the hypothesis is true. The Bayesian approach was deemed to be especially important when the $p$-value yields results that are inconsistent with current knowledge. Nonetheless,~\cite{browner-1987} did not give the explicit formula for diagnostic screening.

For an alternative derivation of the formula, we can let $+$ denote the condition $\ud(\boldsymbol{X})\ge\ud({\boldsymbol{x}})$ and use $-$ to denote a negation of this condition. Applying the laws of probability yields
\begin{equation}
\mbox{Pr}(H_0|+) = \frac{\mbox{Pr}(+|H_0)\mbox{Pr}(H_0)}{\mbox{Pr}(+|H_1)\mbox{Pr}(H_1)+\mbox{Pr}(+|H_0)\mbox{Pr}(H_0)},
\label{eqn:fpr2}
\end{equation}
under the assumption that $H_1$ is a negation of $H_0$. It follows that $\mbox{Pr}(H_1)=1-\mbox{Pr}(H_0)$. It is nonetheless not always the case that $H_1$ is a negation of $H_0$. In the case where $H_1$ is a negation of the null $H_0$, we have $\mbox{Pr}(H_1)=1-\varphi$, where, as noted earlier, $\mbox{Pr}(H_0)=\varphi$. By implication,~\cite{ioannidis-2005} takes the type II error probability to be $\mbox{Pr}(-|H_1)=\beta$ and calls $1-\beta$ the power. Strictly speaking, when $(1-\beta)$ is power, the parameter $\beta$ should be conditioned on a point in the rejection region and not the whole rejection region. In general, a point alternative cannot  be a negation of the null hypothesis. Even though a treatment of power that ignores dependence on effect size is problematic, the diagnostic screening formula appears to inherently ignore this point.  Moreover, the diagnostic screening formula assumes that the power remains the same across different alternatives, which cannot be the case in practical examples, even if one only considers hypotheses in a specific field.

If one makes the additional assignment $\mbox{Pr}(+|H_0)=\alpha$, it follows that
\begin{equation}
\mbox{Pr}(H_0|+) = \frac{\alpha\varphi}{\alpha\varphi+(1-\beta)(1-\varphi)}.
\label{eqn:fpr3}
\end{equation}
$\mbox{Pr}(H_0|+)$ is indeed the false positive rate (or false finding rate~\cite[e. g.][]{mayo-2018}) and it is 1 minus the {\it positive predictive value} (PPV) introduced by~\cite{browner-1987} and popularised by~\cite{ioannidis-2005}. The term used by~\cite{browner-1987} was the {\it predictive value of a positive diagnostic test}. Whereas~\cite{benjamin-2017} (On page 2) take $\varphi$ to be the proportion of null hypothesis that are true,~\cite{colquhoun-2014} calls $\mbox{Pr}(H_1)$ the prevalence rate whilst~\cite{ioannidis-2005} calls it the proportion of true effects. Whilst $\mbox{Pr}(H_0|+)$ is the false positive rate, $\mbox{Pr}(+|H_0)$ is the probability of a false positive under the null hypothesis; the $p$-value. Fisherian tests focus on assessing whether or not the $p$-value is less than or equal to the threshold $\alpha$. In their simulations,~\cite{benjamin-2017} take the $p$-value to be equal to this threshold.

\begin{figure}
\centering
\hbox{
\includegraphics[height=5cm,width=6cm]{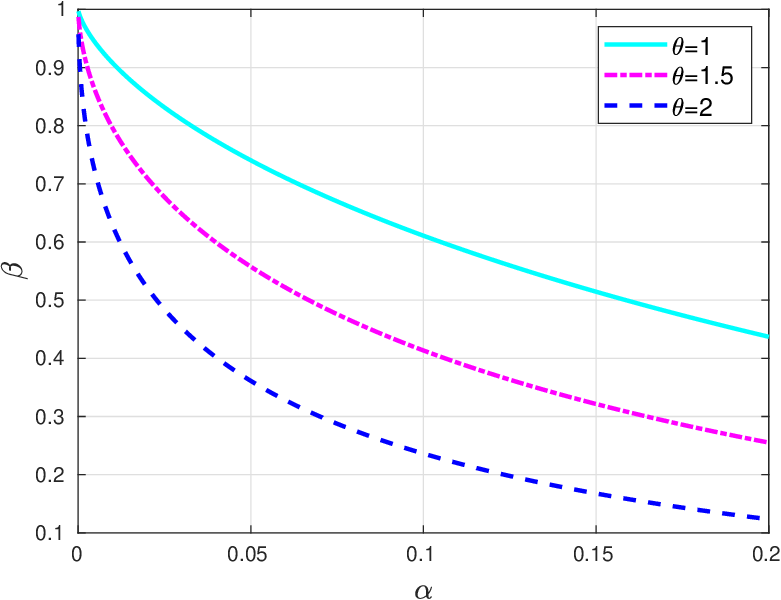}
\includegraphics[height=5cm,width=6cm]{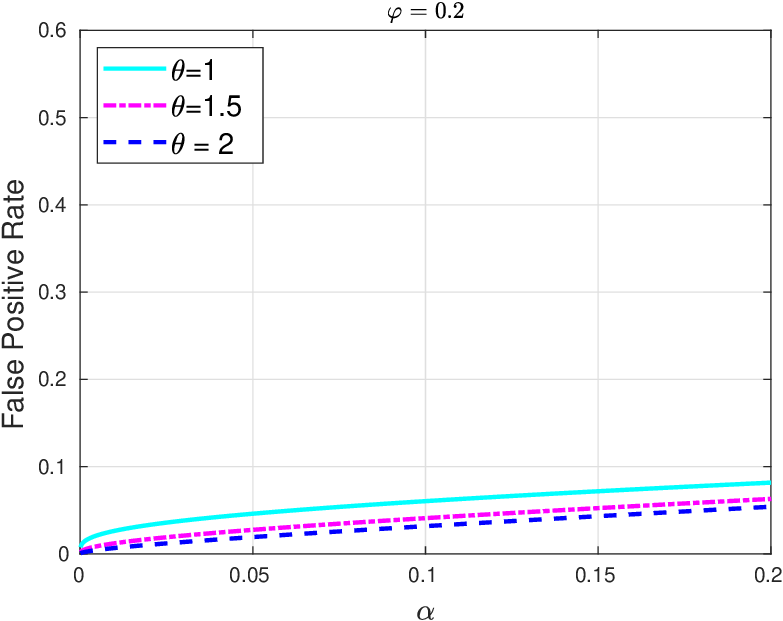}
}
\hbox{
\includegraphics[height=5cm,width=6cm]{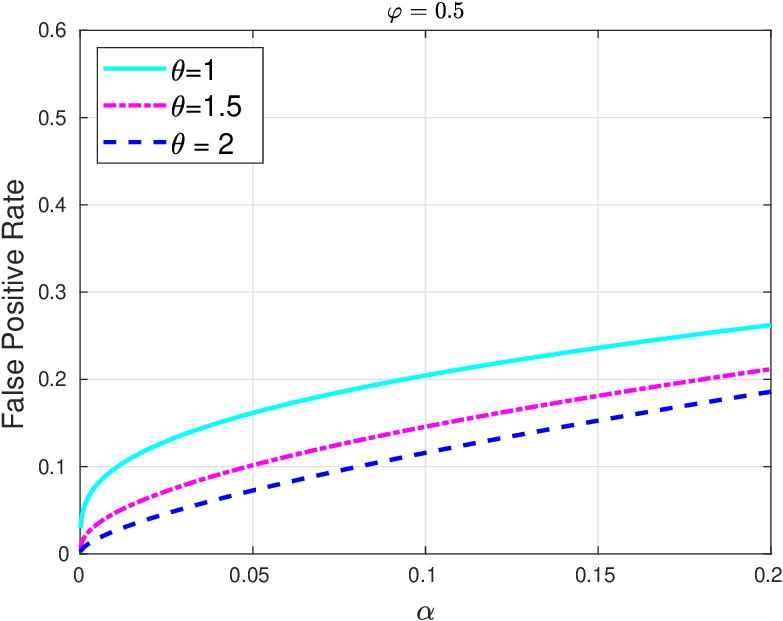}
\includegraphics[height=5cm,width=6cm]{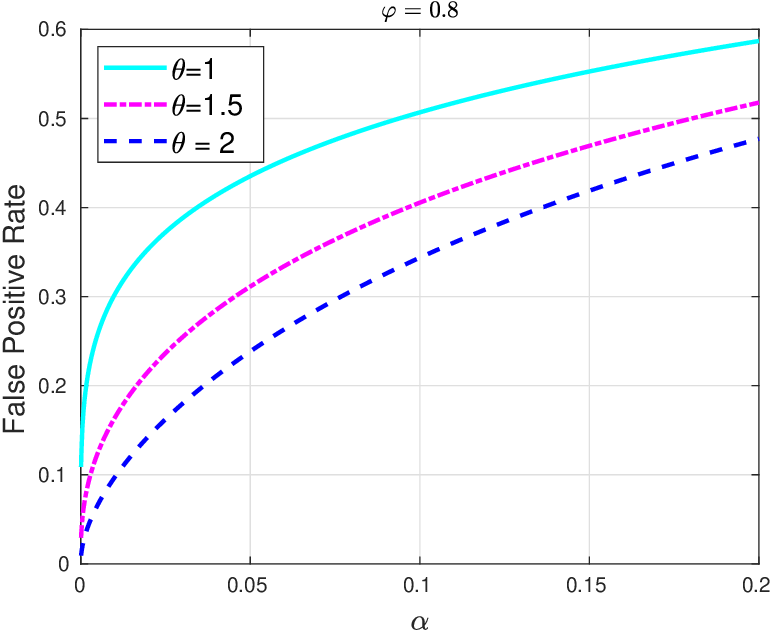}
}
\caption{\it\small Graphs of Type II error probability against Type I error probability (top left) and corresponding graphs of false positive rate versus Type I error probability for different prior probabilities, where $\theta$ is the effect size at which $\beta$ was computed. The values of $\beta$ used to compute the false positive rate are those used to produce the $(\alpha,\beta)$ graphs on the top left.}
\label{fig:pdfs}
\end{figure}

It should be noted that $\mbox{Pr}(H_0|+)$ is a posterior probability of the null. Thus, calls for reform that use the diagnostic screening formula are based on the Bayesian paradigm that is plagued by dependence on the prior $\mbox{Pr}(H_0)$. Although~\cite{colquhoun-2014} casts a veiled aspersion on the notion that this is a Bayesian formula after doing a few calculations where $\mbox{Pr}(H_0)$ is known, this quantity is impossible to compute in practical situations. Furthermore, as noted by~\cite{mayo-2018}, the Diagnostic Screening model mixes up the probability of a Type I error with the posterior probability, thereby introducing confusion. In frequentist tests, reducing the Type I error probability results in increasing the Type II error probability, which is a trade-off.~\cite{mayo-2018} notes that this trade-off disappears in the Diagnostic Screening model: reducing the Type I error rate also reduces the false finding rate. In Fig~\ref{fig:pdfs} on the top left are shown graphs that depict how $\beta$ varies with $\alpha$ for different effect sizes, under the assumption that the sampling distribution is normally distributed. The relationship is clearly nonlinear and as $\alpha$ is reduced, $\beta$ increases.

\subsection{Analysis of Threshold Reduction}
\label{sec:analysis}
In this subsection, an analysis of a more stringent threshold is presented. The suggestion to use a more stringent threshold might re-ignite conflict between two schools of thought, which are a Bayesian account versus an error statistical account (or frequentist approach). There is no agreement within the statistical community on which one is preferable. Adjusting the significance level involves a trade off between type I and Type II errors. Reducing the statistical significance threshold from 0.05 to 0.005 will inevitably increase the probability of Type II errors. Therefore, the cost of increasing the replication rate via a stringent statistical significance threshold is an increase in studies that will unfairly never be allowed to see the spotlight. This effect needs to be unearthed to quantify the extent of the trade off.~\cite{benjamin-2017} do not quantify the extent of consequent Type II errors to counter balance the laudable effects of the more stringent threshold. According to Fig.~\ref{fig:pdfs}, the overall effect is that false positives will be reduced in spite of an increase in the Type II error probability. This is great news for the reform movement.

Another criticism of $p\approx 0.05$ is that it is comparable to Bayes factor of 2.5 to 3.4, even though, unlike the Bayes factor, the $p$-value does not depend on the alternative hypothesis~\citep{benjamin-2017}. Such values of the Bayes factor are considered to be weak evidence for $H_1$~\citep{benjamin-2017}. This is tantamount to another age-old argument that the $p$-value exaggerates evidence against the null hypothesis~\citep{goodman-1992}. After a detailed criticism of the $p$-value,~\cite{goodman-1999} went further to propose the use of Bayes Factors as a better alternative~\citep{goodman2-1999}. As our first response to this, we highlight that significance tests are not comparative measures. The philosophy behind $p$-values is about controlling error probabilities as opposed to estimating posterior probabilities of the null and the alternative~\citep{mayo-2018,mayo-2022}. Significance tests provide a peace-meal approach to assessing the correctness of hypotheses. For a further discussion of differences between posterior probabilities and error probabilities, see~\cite{mayo-2022}. The idea that significance tests exaggerate the evidence can be understood in terms of the {\it Jeffreys-Lindley paradox}~\citep[see][]{mayo-2018,mayo-2022}. The posterior gives a higher probability to the point null hypothesis than would be obtained from the $p$-value because a lot of prior mass is lumped onto the null, the so called spike prior. Otherwise using diffuse priors would give results that agree with significance tests~\citep{mayo-2022}. Spike priors are associated with two tailed significance tests.

Finally, inspite of its highlighted weaknesses, let us consider formula~(\ref{eqn:fpr}), of the false positive rate (also called the {\it false finding rate}~\citep{mayo-2018}). Thinking of the false positive rate as a function of $\alpha$ and $\beta$, denoted by $G(\alpha,\beta)=\mbox{Pr}(H_0|+)$, then the first partial derivatives of G with respect to $\alpha$  and $\beta$ are respectively
\begin{equation}
G_{\alpha}(\alpha,\beta) = \frac{\varphi(1-\beta)(1-\varphi)}{[\alpha\varphi+(1-\beta)(1-\varphi)]^2},
\label{eqn:gprime}
\end{equation}
and
\begin{equation}
G_{\beta}(\alpha,\beta) = \frac{\alpha\varphi(1-\varphi)}{[\alpha\varphi+(1-\beta)(1-\varphi)]^2},
\label{eqn:gprimebeta}
\end{equation}
which are both positive as long as $\varphi\in(0,1)$ and $\beta\in(0,1)$. Thus, $G(\alpha,\beta)$ is an increasing function of each of $\alpha$ and $\beta$. It follows that lowering $\alpha$ should reduce the false positive rate. Likewise, increasing $\beta$ will increase the false positive rate. Since reducing $\alpha$ results in an increase in $\beta$, one may wonder what the combined effect on the false positive rate is. To grasp this, we can consider graphs of the false positive rate versus $\alpha$ that are shown in Fig~\ref{fig:pdfs}. The different graphs correspond to the different prior probabilities of the null (or the prevalence rate of the effect size of interest). It is evident from the graphs that reducing the $\alpha$ leads to a fall in the false positive rate inspite of a corresponding increase in $\beta$. Furthermore, the graphs show substantial sensitivity to the prevalence rate.

Nonetheless finding $G$ empirically remains a challenge, if not altogether impossible, because finding the prior is an elusive exercise.  To make empirical estimates of the false positive rate, one could compute the ratio of false positives to the total number of positives identified by the tests. But this is not possible because we cannot definitively identify false (and true) positives. This may explain why~\cite{benjamin-2017} made hypothetical simulations with different prior ratios, even though they considered only the cases where $\mbox{Pr}(H_0)\ge 5\mbox{Pr}(H_1)$. Making simulations for a specific field or journal requires one to use the respective prior odds ratio. The challenge then is to compute the proportion of hypotheses that are true, and~\cite{ioannidis-2005} suggested that a lot of effort needs to be invested in estimating this quantity.~\cite{benjamin-2017} reported the prior odds ratio to be 1:10 in Psychology, but it is not clear how the estimate was obtained.~\cite{ioannidis-2005} denoted the prior odds ratio by $R$, i. e. $R = (1-\varphi)/\varphi$. Using this notation, one can re-write the false positive rate as
\begin{equation}
\mbox{Pr}(H_0|+) = \frac{\alpha}{\alpha+(1-\beta)R}.
\label{eqn:fpr4}
\end{equation}
 The false positive rate is less than half if $R\ge\alpha/(1-\beta)$.  

We understand and appreciate the desire to bolster replication rates in science. Benjamin et al.~\citep{benjamin-2017} highlighted that, in psychology and experimental economics, replication rates for studies with $p<0.005$ were double those obtained with $0.005<p<0.05$. This is arguably the main motivation for reducing the significance threshold down to 0.005. It was not reported what the replication threshold was in each case, but it seems the same threshold of 0.05 was used to determine replication in each case. One can think of the doubling of replication rates due to reducing the significance threshold to be tantamount to doubling the probability of true positives, $\mbox{Pr}(H_1|+)$, what~\cite{browner-1987} and~\cite{ioannidis-2005} termed the positive predictive value. This begs the question: How does changing the significance level $\alpha$ by some factor (say $r$) affect the replication rate? Put another way: By what factor should the significance level be reduced in order to increase the replication rate $n$-fold? In order to address this question, let $\kappa=(1-\beta)R$ and let $\gamma=\mbox{Pr}(H_1|+)$. Thinking of $\alpha$ as a function of $\gamma$, $\psi(\gamma)$, upon rearranging~(\ref{eqn:fpr4}) and using the foregoing substitutions, one obtains
\begin{equation}
\psi(\gamma)=\frac{\kappa(1-\gamma)}{\gamma}.
\end{equation}
Our interest is to assess
\begin{equation}
r = \frac{\psi(\gamma)}{\psi(n\gamma)}=\frac{n(1-\gamma)}{(1-n\gamma)}.
\label{eqn:factor}
\end{equation}
Doubling the replication rate corresponds to $n=2$, which is possible provided $\gamma<\frac{1}{2}$. Thus the true positive rate must be less than half for it to be possible to double it by simply reducing the significance threshold. More generally, the replication rate can be increased $n$-fold by reducing the significance threshold provided $\gamma<1/n$. In particular, reducing the significance threshold by a factor of 10 (as proposed by~\cite{benjamin-2017}) will result in the replication rate increasing $n$-fold provided the true positive rate $\gamma = \frac{(10-n)}{9n}$. Thus reducing the threshold by a factor of 10 will double the replication rate (i. e. $n=2$) provided the true positive rate $\gamma=4/9$ (also termed the positive predictive value). The point of the foregoing analysis is that whether or not a specific tinkering with the significance threshold will result in a desired increase in the replication rate depends on the true positive rate at the current (or prevailing) significance threshold.

As far as we understand, the only available tools for identifying true (and false) positives are the error statistics under criticism. Take an instance where tests are done to determine if a malaria treatment is effective. The effectiveness (or none-effectiveness) of the treatment can only be determined by performing statistical tests via error probabilities. Any future refutations will depend on error statistics, which are probabilistic in nature. Therefore, it will be absurd to talk of false positives in a definite sense, especially when the significance threshold used is one deemed questionable. The tinkering with the significance threshold will not be a magic wand for the big problems that confront scientific exploration, such as optional stopping rules, data dredging e.t.c. We agree with~\cite{senn-2002} that ''replication probabilities are not of direct relevance to inferential meaning`` and $p$-values should be accepted for what they are. If honesty and integrity can be restored in science, error probabilities will be effective tools in promoting true discoveries. Whereas proponents of reform tend to want error statistical methods to be evaluated by Bayes Factors, the use of Bayes Factors and like alternatives will not provide solutions to the aforementioned problems.
\subsection{Other Reforms and the Error Statistician}
\label{sec:others}
This section discusses a number of recommended reforms mentioned in the introduction.~\cite{amrhein-2019} argued for the use of confidence intervals instead of significance tests, termed compatibility intervals in~\cite{amrhein-2019b}. Other calls for the use of confidence intervals include~\cite{anderson-2019}, \cite{colquhoun-2019} and~\cite{greenland-2019}. These calls are very welcome and it is noteworthy that parameter estimates are often reported with associated confidence intervals. This is fine because there is a duality between error probability tests and confidence intervals. Should confidence intervals be done at the expense of $p$-values? Not really because the two can complement each other to aid decision making. 

Meanwhile, there are recommendations to provide $p$-values in a continuous way~\citep{amrhein-2019b,mcshane-2019}. What are their implications? Such recommendations are tantamount to retiring significance tests altogether.~\cite{greenland-2019} suggested that $p$-values should be reported as equality rather than saying $P<\alpha$, recommending doing away with thresholding altogether. Proponents of reporting $p$-values in a continuous way argue that such values can still be used as evidence for or against the null hypothesis without a threshold. Such an approach will, however, amount to an undisclosed threshold. A decision between the null and alternative hypothesis is a dichotomous decision, whether one discloses the threshold or not. How can one abandon thresholding and still use $p$-values? Furthermore, how can one abandon thresholding and still use confidence intervals? These are rhetorical questions. Moreover, confidence intervals cannot be computed without using some threshold. 

There are many situations where dichotomous decisions are required and it may not be wise to merely give decision makers confidence intervals in such situations. Scientists should be trusted to determine when to provide confidence intervals along with the test results. It is noteworthy that Fisher's work on hypothesis testing also provided what was then referred to as fiducial intervals~\citep[see][]{fisher-1955}, albeit marred by the sin of probabilistic instanciation. At this point, it should be noted the ASA's statement on $p$-values is not against thresholding. In particular, Principle 3 says that ''decisions should not be based {\it only} on whether a $p$-value passes a specific threshold``~\citep{wasserstein-2016}. Whilst it is true that statistical significance does not necessarily equate to practical or scientific significance, a threshold may be selected to reflect practical or scientific significance. A real world costs and benefits analysis may be used to inform the choice of threshold.
\subsubsection{Decision Analysis}
~\cite{manski-2019} advocated for supplanting hypothesis testing with statistical decision theory. An earlier call to use Bayesian decision theory was echoed by~\cite{bernardo-2002} who suggested using an information criterion to make a decision between the null and alternative hypothesis. The statistical decision theory presented in~\cite{manski-2019} is a Bayesian version of expected utility theory. The weaknesses of expected utility theory have been  documented~\cite[e.g][]{kah-79}. It is good for scientists to have a choice of theories, but it should be left to them to make a decision based on their circumstances, and expected utility theory is not necessarily better than, and need not supplant, significance testing. The formulation presented by~\cite{manski-2019} is a generalisation of Bayesian concepts presented by~\cite{wonnacott-1986} with a focus on trial data statistics. Moreover, Bayesian hypothesis testing is performed using the posterior probability to make the decision and not the $p$-value of classical hypothesis testing.

\begin{figure}
\centering
\hbox{
\includegraphics[height=6cm,width=6cm]{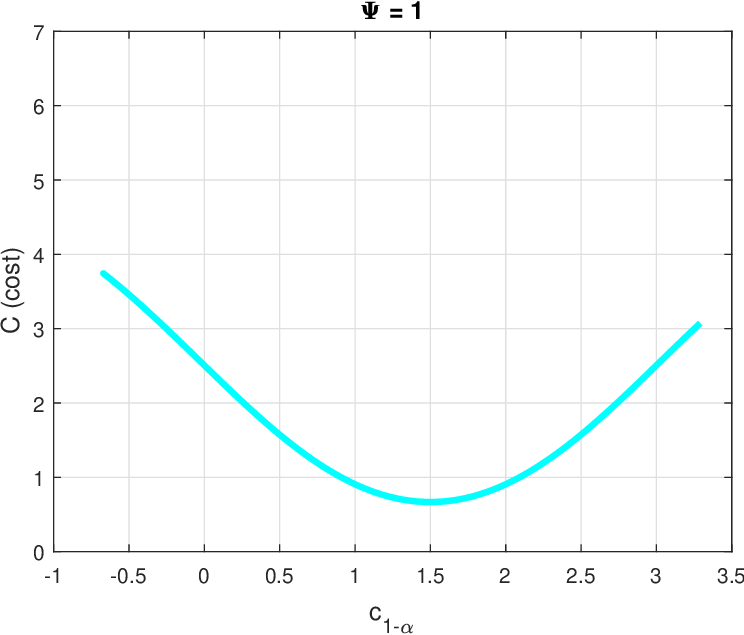}
\hspace{0.4cm}
\includegraphics[height=6cm,width=6cm]{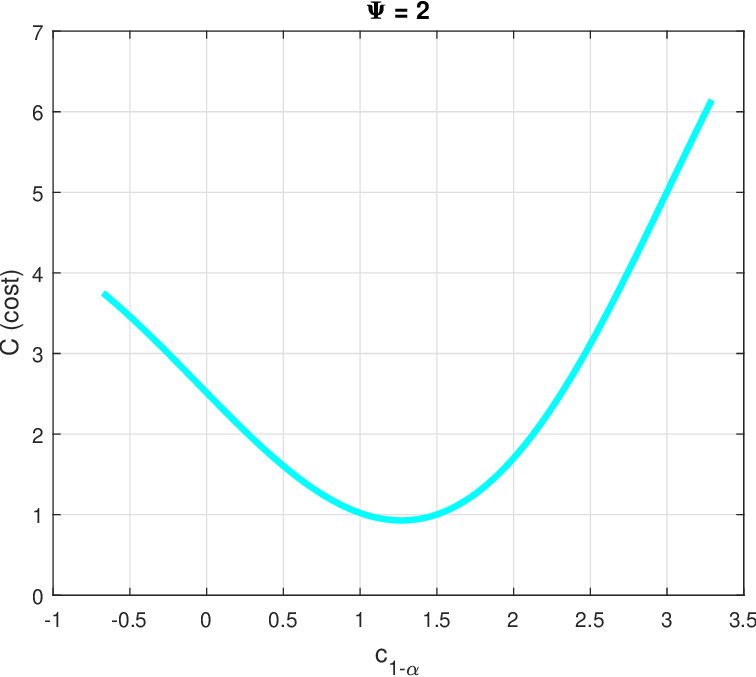}}
\vspace{0.4cm}
\hbox{
\includegraphics[height=6cm,width=6cm]{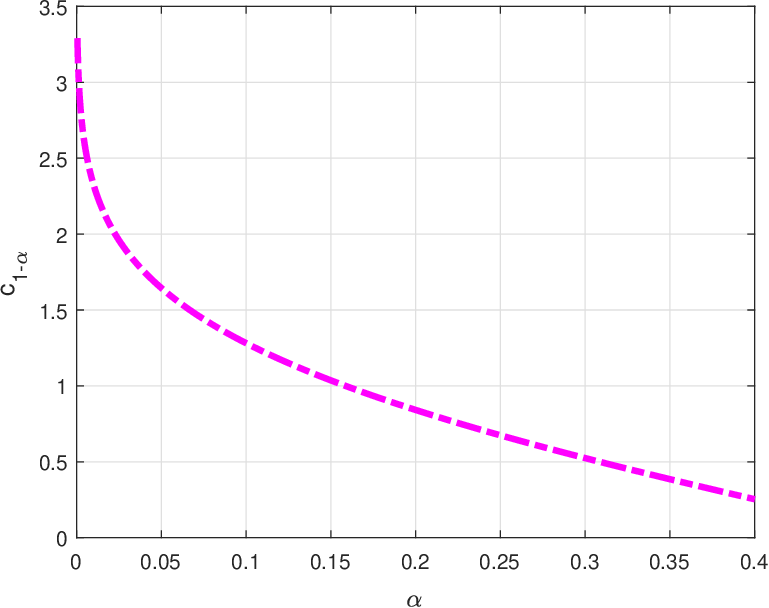}
\hspace{0.4cm}
\includegraphics[height=6cm,width=6cm]{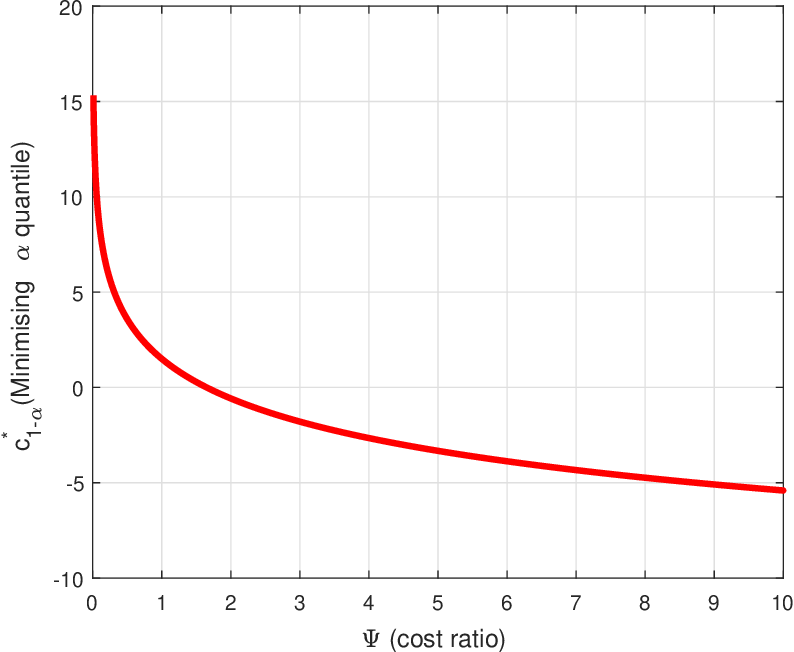}
}
\caption{\it\small Top graphs show the expected cost as a function of the critical value for two cost ratios. The bottom left is the a graph of the critical value as a function of the significance level whilst the bottom righ is a graph of the minimising critical value (or $\alpha$-quantile) as function of the cost ratio. These graphs are obtained under the assumption that test statistic is normally distributed.}
\label{fig:costs}
\end{figure}

\cite{wonnacott-1986} argued that a threshold for classical significance testing can be chosen in a Bayesian framework to reflect potential losses in the event of a type I or type II error. Here, we demostrate that this can be done within the typical frequentist framework. Consider a situation of a factory that produces devices for sale. Through some sampling approach, it may be found that a proportion $\varphi$ of the devices are good and $(1-\varphi)$ are faulty. Let the cost of identifying a device as faulty when it is good (type I error) to be $P_0$ and the cost of identifying a device as good when it is faulty (type II error) to be $P_1$. The threshold $\alpha$ (equivalently $c_{\alpha}$ and $c_{1-\alpha}$) can be chosen to minimise the expected cost
\begin{equation}
 C(c_{1-\alpha})=\varphi(1-F_0(c_{1-\alpha})) P_0 +(1-\varphi)F_1(c_{1-\alpha}) P_1
\end{equation}
where $\alpha=\int_{c_{1-\alpha}}^\infty f_0(x)\ud x=1-F_0(c_{1-\alpha})$ is the probability of a type I error, $\beta = \int_{-\infty}^{c_{1-\alpha}}f_1(x)\ud x=F_1(c_{1-\alpha})$ is the probability of a type II error and $f_0(x)$ and $f_1(x)$ are probability density functions of the test statistic under the null and the alternative hypothesis respectively. It follows that

\begin{equation}
 C'(c_{1-\alpha})=-\varphi P_0 f_0(c_{1-\alpha})+(1-\varphi)P_1f_1(c_{1-\alpha})
\end{equation}
At the extrema, $C'(c_{1-\alpha})=0$. If the test statistic is normally distributed with variance $\sigma^2$ and mean $\mu_0$ under the null hypothesis and $\mu_1$ under the alternative hypothesis, the minimum is attained when 
\begin{equation}
 c_{1-\alpha}=\frac{\sigma^2}{(\mu_0-\mu_1)}\log\left[\frac{(1-\varphi)P_1}{\varphi P_0}\right]+\frac{1}{2}(\mu_0+\mu_1)
\end{equation}
A second derivative test will establish that this is a minimiser provided $\mu_0<\mu_1$, regardless of the prior odds ratio nor the cost ratio $\Psi=P_1/P_0$. Graphs of the expected cost as a function of the critical value for two cost ratios are shown in Figure~\ref{fig:costs}. Evidently, the minising critical value decreases with an increasing cost ratio as shown in the bottom right graph shown in Figure~\ref{fig:costs}.

What are the cost implications of reducing the significance threshold? Firstly, note that the minimiser will not necessarily coincide with a selected significance threshold. Reducing the significance threshold will increase the critical value, $c_{1-\alpha}$. The net effect on the cost will depend on the cost ratio $\Psi$ and the corresponding power, $1-\beta$. The cost is an increasing function of $\alpha$ if $\Psi\frac{(1-\varphi)}{\varphi}<\frac{f_0(c_{1-\alpha})}{f_1(c_{1-\alpha})}$ and decreasing otherwise. A higher value of the cost ratio, $\Psi$, leads to a lower minimising critical value. The resulting expected cost at the lower minimiser will be higher. A potential increase in the expected cost when the critical value is increased is an indictment against reducing the significance threshold. On the other hand, a potential reduction in the expected cost when the critical value is increased incentivises reducing the significance threshold. In both cases shown in Figure~\ref{fig:costs}, the minimising critical value is less that 1.5 ($c_{1-\alpha}^*<1.5$). Meanwhile, $c_{1-\alpha}<1.5$ corresponds to $\alpha>0.07$, values rarely considered as thresholds in significance testing.

In concluding the cost function analysis, let us consider two practical scenarios that pose a challenge to it. In the first scenario, one can consider the case of determining if there is global warming or not. Costs associated with falsely concluding that there is global warming are intractable, and will differ with the diverse decision scenarios contingent on the presence or absence of global warming. Likewise, one can consider the case of drug development where the efficacy of a new drug is assessed through clinical trials. It seems reasonable to aim to minimise the false positive rate rather than quantifying the costs of errorneous positives. In a nutshell, while cost function analysis may appear attractive, there are many practical situations that render its application impractical or of limited value.  In situations where cost function analysis is warranted, significance testing may be done using minimising thresholds.
\subsubsection{Severity Formulation}
Note that Fisherian and N-P tests are used to control the probabilities of erroneously concluding that the available data provide evidence for or against hypotheses under consideration. One who employs these piecemeal tools to make a decision regarding competing hypotheses is an {\it error-statistician}~\citep{mayo-2018}. It is not enough to have a hypothesis agreeing with the data by passing a given test. According to~\cite{mayo-2018}, it has to satisfy another condition as part of what is termed the {\it severity} requirement, captured by the following definition:
\begin{definition}\label{de1}
(Strong Severity) A hypothesis $H$ passes a test with severity provided
\begin{itemize}
\item[S-1:] $H$ agrees with the data $\boldsymbol{x}$.
\item[S-2:] With a high probability, $H$ would not have passed the test if it were false.
\end{itemize}
\end{definition}
When this definition is satisfied, then a hypothesis or claim is corroborated by the data.  Weak severity, on the other hand, denies that $H$ is warranted if the test would probably have passed $H$ even if it were false. In the Fisherian and N-P tests, the function $\ud(\boldsymbol{X})$ reflects how well or poorly the data accord with the hypothesis $H_0$. The distribution of $\ud(\boldsymbol{X})$ is used to characterise the probative abilities of the underlying inferential rule for the task of unearthing flaws and mis-representations of the data.  Only the error statistician can be a severe tester; the severity requirement entails, but is not limited to, computing error probabilities. When a hypothesis passes or fails a significance test, the severe tester still probes further to determine which set of values within the relevant region are warranted. This can address the concerns of~\cite{amrhein-2019} who argued for using confidence intervals. In this way, a dichotomous decision can co-exist with confidence intervals. It is indeed common to find values of interest reported with their corresponding standard errors. Results of significance tests are usually reported alongside such values. 

To make severity assessments,~\cite{mayo-2018} defines the severity function as the severity with which claim $C$ passes test $T$ with data $\boldsymbol{x}$:
\begin{equation}
\mbox{SEV}(T,\boldsymbol{x},C) = \mbox{Pr}(\mbox{worse fit with C};\sim C).
\end{equation} 
In the above formula, $\sim C$ denotes the negation (or denial) of claim $C$. The severity function is what the severe tester can use to probe the parameter space further. Severity improves upon confidence intervals by giving them an inference interpretation and by giving a severity assessment associated with different hypotheses. In Section~\ref{sec:examples} we give an example to illustrate the application of the concepts: Fisherian tests, N-P tests and severity tests in real problems. The next section discusses replication and  distributions of $p$-values, and if there is a role they can play to judge the strength of evidence.
\section{Reproducibility and $p$-Value Distributions}
\label{sec:pvalues}
Re-echoing the words of~\cite{amrhein-2019b}, \cite{gibson-2021} argued that there is no replication crisis in science if we do not expect it. He appealed to the works of~\cite{hung-1997} and \cite{lambert-1982} for the distribution of $p$-values under the alternative hypothesis. A brief review of the distribution of $p$-values for a single sample test will be given here, including implications to replication in science. In particular, the extent to which such a distribution can be used to quantify the strength of evidence is discussed here.

Since the $p$-value is a quantity obtained from the test statistic that is a random variable, it can also be considered to be a random variable. Under the null hypothesis, the $p$-value follows the standard uniform distribution, $U[0,1]$, regardless of the distribution of the test statistic. On the other hand, if the test statistic is normally distributed, the distribution of $p$-values under the alternative hypothesis is asymptotically log-normal and varies with the test statistic~\citep{lambert-1982}. For this discussion, let $X$ be the random variable of interest that has mean $\mu$ and variance $\sigma^2$ and let $x_n$ be the corresponding sample mean from $n$ observations. Setting $\delta=\mu/\sigma$, the probability density function of the $p$-value is~\citep{hung-1997}
\begin{equation}
g_{\delta}(p)=\phi(c_p-\sqrt{n}\delta)/\phi(c_p),\quad 0<p<1,
\end{equation}
where $c_p$ is the $(1-p)$th percentile of the standard Gaussian random variable and $\phi$ is the Gaussian density function. The corresponding distribution function is
\begin{equation}
G_\delta(p)=\int_0^p g_\delta(s)\ud s = 1 -\Phi(c_p-\sqrt{n}\delta).
\label{eqn:df}
\end{equation}
These are for a one sample 1-tailed test. 

\cite{gibson-2021} notes that the distribution function of $p$ can be used to judge the strength of evidence. This distribution depends on the sample size $n$ and the effect size $\delta$. Therein, it is argued that, for a given value of $p$, a higher value of $G_\delta(p)$ signifies stronger evidence in favour of $p$. This need not be, however, because the prevailing scenario is not necessarily that of the alternative hypothesis.

The distribution function is increasingly steep with increasing sample size $n$ and effect size $\delta$~\citep{hung-1997}. For an anticipated magnitude, $\mu=\mu^*$, the sample size chosen to maintain power $1-\beta$ can be set using the relation
\begin{equation}
n = \{\sigma(c_\alpha+c_\beta)/{\mu^*}\}^2.
\end{equation}
When the power increases, the sample size required to maintain the higher power increases. Furthermore, reducing the $p$-threshold requires increasing the sample size to maintain the same power. For a fixed effect size $\delta$, the power is given by~\citep{hung-1997}
\begin{eqnarray*}
\mbox{Pr}(\ud(\boldsymbol{X})>c_\alpha|\delta) &=&\mbox{Pr}(P<\alpha|\delta)\\
&=&G_\delta(\alpha).
\end{eqnarray*}
When the effect size $\delta$ corresponds to an estimate from observations, \cite{gibson-2021} calls $\mbox{Pr}(\ud(\boldsymbol{X})>c_\alpha|\delta)$ the reproducibility probability, albeit considering the form for a two sample, 2-tailed test with common mean. One can modify the reproducibility probability expressions given in~\cite{gibson-2021} and~\cite{goodman-1992} to obtain the reproducibility probability result for a one sample, 2-tailed test as
 \begin{equation}
 F_\delta(\alpha)=1-\Phi(c_{\alpha/2}-\sqrt{n}\delta_o)+\Phi(-c_{\alpha/2}-\sqrt{n}\delta_o),
 \label{eqn:rp2}
 \end{equation}  
 where $\delta_o$ is the observed effect size. The corresponding observed test statistic is $d_o=\sqrt{n}\delta_o$. For an observed $p$-value, $p_o$, $d_o = -\Phi^{-1}(p_o/2)$~\citep{gibson-2021}.  Recall that earlier we discussed the false positive rate $\mbox{Pr}(H_0|+)$. The reproducibility probability differs from $1-\mbox{Pr}(H_0|+)$, the true positive rate. Furthermore, the $\beta$ in~(\ref{eqn:fpr4}) is not quite the type II error probability because it is not conditioned on a specific effect size. Also note that the false positive rate is estimated from a family of experiments within a specific field whereas the reproducibility probability is specific to a type of experiment and test statistic.~\cite{senn-2002} calls a formula like~(\ref{eqn:df}) and (\ref{eqn:rp2}) the replication probability and provides some insights on its interpretation, arguing that it would be absurd for inference at a given moment to depend on replication at a future time. \cite{gibson-2021} considers  $G_{\delta}(\alpha)$ to be a more direct measure of replication, which is the deliberate repetition of an initial experiment to confirm its findings. Unfortunately, $G_\delta(\alpha)$ may just be a theoretical quantity with no practical bearing because the prevailing scenario is not necessarily that of the alternative hypothesis. Moreover, the effect size in the reproducibility probability is specific to the observations and does not cover the whole alternative region; but the probability $\mbox{Pr}(+|H_1)$ behaves like power because $\mbox{Pr}(+|H_1)=\int_{\Omega_1}\mbox{Pr}(+|\delta)\ud\delta/\int_{\Omega_1}\ud\delta$, where $\Omega_1$ is the region corresponding to the alternative hypothesis.
 
 Meanwhile, there have been calls for pre-publication confirmatory studies before strong conclusions can be drawn about the validity of the alternative hypothesis~\citep{gelman-2014,nosek-2012}. This is definitely useful in tackling the issue of multiple comparisons (consequently inadvertent $p$-hacking) highlighted by~\cite{gelman-2014}. A testimony of two witnesses is surely stronger than that of one. If one is thinking of sampling the distribution of $p$-values, a two point sample is still quite small to give a glimpse of the underlying distribution, whilst performing many experiments is unrealistic for many practical purposes. Nonetheless further experiments can still refute conclusions drawn by pre-publication confirmatory studies. If one is interested in the empirical distribution of $p$-values, bootstrap approaches could be summoned for assistance. Otherwise, for a given data set (or experiment), the computed $p$-value should be used to judge whether or not the alternative hypothesis is warranted without entangling oneself with questions of replication.
\section{African Savanna Fires}
\label{sec:examples}
Here we consider an example of the autocorrelation of the proportion of burnt area of Savanna wildfires over sub-Saharan Africa, which was used by~\cite{machete-2023} to detect oscillations of fire patterns. The data used is the annual proportion of burnt area in Savanna regions over the time period from 2001 to 2020. Therein, it was concluded that wildfires exhibit cycles of 5 years. One would, therefore, expect significant autocorrelation at a time lag of 5 years. The autocorrelation $\rho$ corresponding to a time lag of 5 years was found to be 0.3839. The $p$-value can be computed using the $t$ test, where $t$ is the statistic that follows Student $t$'s distribution. Given a correlation coefficient $\rho$ and an observation vector of length $n$, the relevant statistic is given by
\begin{equation*}
t =\frac{\rho\sqrt{n-2}}{\sqrt{1-\rho^2}}.
\end{equation*}
It follows a $t$ distribution with $n-2$ degrees of freedom.

Let $Y_t$ be the annual proportion of Savanna burnt area over sub-Saharan Africa. Instead of the auto correlation function, we can consider the linear model
\begin{equation}
Y_t = \beta_0 + \beta_1 Y_{t-\tau}+\varepsilon_t.
\label{eqn:lag}
\end{equation}
where $\beta_0$ and $\beta_1$ are constants and $\varepsilon_t$ is a random error term. If we let $\hat{\beta}_1$ be the random variable that yields estimates of $\beta_1$, then the statistic of interest evaluated at the data set is $d = \frac{\mathbb{E}[\hat{\beta}_1]-\beta_1}{\sigma_{\hat{\beta}_1}}$. We are interested in testing the null hypothesis $H_0: \beta_1 = 0$ against the alternative hypothesis $H_1: \beta_1\neq 0$. The statistic for performing this test is 
\begin{equation*}
d = \frac{\mathbb{E}[\hat{\beta}_1]}{\mbox{s.e.}(\hat{\beta_1})}=\frac{r\sqrt{n-2}}{\sqrt{1-r^2}},
\end{equation*}
where $r$ is an estimate of the correlation coefficient. 

 The data yielded estimates of $\mathbb{E}[\hat{\beta}_1]=0.5782$ and $\mbox{s.e.}(\hat{\beta}_1)=0.1654$ with a $p$-value of $2.37\times 10^{-4}$. Thus the null hypothesis is rejected in favour of the alternative. Here, $\hat{\beta}_1$ is a random variable. The $5\%$ lower confidence limit for $\beta_1$ is 0.30529. This follows from the duality between confidence intervals and significance tests. To bring in severity construals to confidence limits, we can consider the hypothesis (or claim) $h$: $\beta_1>0.30529$. The estimate $\mathbb{E}[\hat{\beta}_1]=0.5782$ accords with the claim $h$, implying that the data accords with $h$. The value $\beta_1=0.30529$ falsifies the claim. Furthermore, values $\hat{\beta_1}<0.5782$ constitute worse fit accorded to the model. The value of the severity function associated with the claim $h$ is $\mbox{SEV}(h)=\mbox{Pr}(\hat{\beta_1}<0.5782;\beta_1=0.30529)=0.95$, which is quite high. Thus, the claim that $\beta_1>0.30529$ is warranted with severity 0.95.
 
 What then is the replication probability under the alternative hypothesis? Let us consider the standard significance level $\alpha=0.05$. In this case $c_{\alpha/2=1.96}$ and the observed test statistic is $d_o=\frac{0.5782}{0.1654}=3.5$. The important quantities to be plugged into~(\ref{eqn:rp2}) are $c_{\alpha/2}-d_o=-1.54$ and $-c_{\alpha/2}-d_o=-5.46$, whence the replication probability is $F_{\delta_o}(\alpha)=1-\Phi(-1.54)+\Phi(-5.46)=0.93$. Hence there is a 93.8\% chance that a future study will replicate the results. However, any future study will have to use the same data used here because these are real world observations that are not from an experiment. 
\section{Discussion}
\label{sec:discussion}
Since their introduction by Fisher and further developments by Neyman and Pearson, error statistical tests have faced controversies which are renewed time and again. This is evidenced by calls for caution by the American Statistical Association~\citep{wasserstein-2016} against mis-interpretations and abuses of $p$-values with an attempt to provide guidance and correct abuses of the tool. The editorial contained the ASA's statement on $p$-values, which statement was meant to curb the replication crisis concerning studies that used $p$-values. The statement was later followed up by a special issue entitled {\it Moving to a world beyond $p<0.05$}~\citep{wasserstein-2019}. This paper considered a number of reforms proposed therein and elsewhere. These include reducing the significance threshold for $p$-values~\citep{benjamin-2017,benjamin-2019} to use a more stringent one, prepublication of confirmatory studies~\citep{nosek-2012,gelman-2014}, abandoning the use of $p$-values altogether in favour of reporting confidence intervals~\citep{amrhein-2019b,amrhein-2019}, reporting $p$-values in a continuous way~\citep{mcshane-2019} and abandoning thresholds of $p$-values~\citep{greenland-2019}, using cost benefit analysis~\citep{mcshane-2019}, and supplanting significance tests with statistical decision theory~\citep{manski-2019}. These calls have their merits, even though some of them might place shackles on scientific exploration, while others are likely to remove the demarcation necessary to distinguish true science from false science.~\cite{benjamin-2017} also contained names of people who signaled their support for the 0.005 threshold instead of the widely used threshold of 0.05. The inclusion of such signatories in a scientific paper is a strange one, especially in a paper that calls for good statistical practice. Such an inclusion inadvertently encourages biased selection.  

The urge to use a more stringent significance thres-hold seeks to address a major concern about replication in science. Nonetheless, the admonition to reduce the significance level fails to address the point that there is a trade off between type I and type II errors. In particular, a more stringent threshold is likely to increase the rejection of findings that should otherwise stand. In persuading their case,~\cite{benjamin-2017} provided calculations done under varying power and fixed significance threshold ($\alpha=0.05$ and $\alpha=0.005$ separately). Varying power is equivalent to varying  the Type II error rate and the trade off with the Type I error rate should be accounted for. This paper has done calculations that took the trade off into account. It was found that reducing the significance threshold with an appropriate Type II error rate still leads to a reduction in the false positive rate. The extent of the reduction in the false positive rate, however, varies greatly with the prior odds ratio. 

A discussion of the distribution of $p$-values under the alternative hypothesis was also presented. Challenges associated with using a  more stringent threshold were documented. In particular, reducing the $p$-value threshold requires increasing the sample size in order to maintain the same power. Furthermore, $p$-value distributions should be interpreted with caution because the prevailing real-world scenario may not necessarily be that of the alternative hypothesis. Deducing replication probabilities from them can thus be misleading. One should rather conduct confirmatory experiments to reduce the risk of being misled by a single $p$-value. Nonetheless we also caution that a sample of two $p$-values is too small to give an indication of the whole distribution. It is, however, practically prohibitive to perform sufficient experiments to sample $p$-value distributions. In light of these concerns, $p$-values should be allowed to play their inferential role without being confounded with replication probabilities. 

One of the arguments given by~\cite{benjamin-2017} for reducing the threshold is that, compared to the Bayes factor, the $p$-value exaggerates  evidence. This is an old argument that plays into the Jeffreys-Lindley paradox~\citep{lindley-2000}, which pinpoints the root of the misconception to placing a lot of prior mass on the point null. For two-sided hypothesis testing,~\cite{shi-2021} have shown that the $p$-value is asymptotically equivalent to transformations of Bayesian posterior probabilities under uninformative prior distributions. Their work was an extension of that of~\cite{casella-1987} who reconciled the $p$ value to Bayesian posterior probabilities for one-sided hypothesis tests. That said, it should be noted that $p$-values are neither posterior probabilities nor inferior to them. This paper has shown that reducing the significance level will necessarily lower the false positive rate regardless of the prior odds ratio. Moreover, the effect of using different prior odds ratios on the benefit of changing the threshold showed massive variations. Since the prior odds ratio depends on estimates of the proportion of hypotheses that are true, its knowledge eludes us because it is well-nigh impossible to accurately know the proportion of null hypotheses that are true. The source of the difficulty emanates in part from challenges in determining the relevant population within which to compute the proportion. The Bayesian paradigm falls prey to this dilemma.

How does reducing the significance level bolster the replication rate? In this paper, it has been shown that, whether or not reducing the significance level will bolster the replication rate depends on the true-positive rate at the prevailing significance threshold. Unfortunately, laying a grip on the true-positive rate is likely to prove a futile exercise. Notwithstanding that, the positive predictive value can be increased $n$-fold provided the true positive rate is less than $1/n$. In particular, this paper has shown that reducing the significance level by a factor of 10 will double the overall replication rate provided the prevailing true-positive rate equals $4/9$.

What about the suggestion to abandon $p$-values in favour of confidence intervals? As for confidence intervals, they are often valuable to provide and there is a duality between confidence intervals and error statistical tests; but they need not be given at the expense of error probabilities. On the contrary, confidence intervals can supplement error probabilities to aid decision making. Severity construals advocated for by~\cite{mayo-2018} provide an alternative way to probe the parameter space to supplement significance tests. It should be understood that confidence intervals do not provide an escape route from thresholds because they are computed based on thresholds. Whilst some have argued that $p$-values should be reported in a continuous way, deciding whether the reported value provides evidence for or against the null hypothesis requires some threshold, even if one does not disclose it. Moreover, reporting a $p$-value cannot be done outside the null and alternative hypothesis dichotomisation because they are computed under the assumption that the null hypothesis holds. The scientific community is, therefore, faced with the task of deciding whether or not to use null hypothesis significance testing as an integral part of the scientific method. Just like confidence intervals, cost benefits analysis can be used to supplement significance testing. At best, the analysis can be an integral part of significance testing, especially in deciding on practically relevant thresholds. Whereas~\cite{manski-2019} suggested supplanting significance testing with statistical decision theory, the theory he put forth is reminiscent of expected utility theory that is fraught with practical challenges. In spite of that, expected utility theory formulation can be tailored for incorporation into classical hypothesis testing. Moreover, it can be shown within expected utility theory that reducing the $p$-value threshold can be suboptimal.

To illustrate the value of error statistical techniques, we considered the proportion of savannas burnt area over sub-Saharan Africa. In the example, the use of severity functions was demonstrated to decide which values to block and which ones are warranted. These supplemented significance test results supplied. Embracing uncertainty need not equate to remaining indecisive, but rather being able to make (and/or guide) decisions under uncertainty. In some cases, the decision is between rival theories where neutrality is not an option. Reporting estimated values with associated confidence intervals can enhance decision making  especially when the computed $p$-value is close to the borderline. We concur with~\cite{amrhein-2019} that $p$-values should not be used blindly, but we do not think they should be retired. As~\cite{cox-1977} stated, ``The object of the significance test is not to supplant the personal judgement of the research worker.'' On the other hand, the point of severity functions SEV is to consider different discrepancies from the null hypothesis and indicate how severely they pass. Severe testers do not use rigid thresholds. In the case of the proportion of savanna burnt area over sub-Saharan Africa, the computed $p$-value was quite small, giving decisive evidence against the null hypothesis. Yet confidence intervals~\footnote{Obtained via severity arguments.} may still provide helpful information for decision making. Scientists should not be slaves to, but masters over tools and thresholds. 

As~\cite{fisher-1955} suggested, the choice of significance level should ideally be left to the scientist, despite using the 0.05 threshold in his farm experiments~\citep{box-1976}. Sometimes errors will be made and false theories will flourish, but this can only be for a time because science is brutally self-correcting. This is evidenced by the so-called replication crisis even within a world of $p<0.05$. It is the 0.05 threshold that has faithfully helped practitioners to identify false discoveries. Credit should be given where it is due. Nonetheless there are no incentives for replication studies. Even though new discoveries may engender efforts for confirmation or refutation, top journals tend to be interested in novelty~\citep{nosek-2012}. In spite of this, it should be established whether or not the associated false-positive rates (equivalently, replication rates) are consistent with the prevailing significance thresholds. Embracing uncertainty includes accepting the inevitable occurrence of false-positives and critically assessing whether these are consistent with $p$-value thresholds. There are surely other problems in science that affect replication rates such as data dredging, optional stopping and P-hacking. The ``redefine significance'' authors admit that lowering the $p$-value will not correct these more serious problems. 

Abandoning significance thresholds is tantamount to abandoning the use of $p$-values altogether. We are, therefore, at the crossroads to decide whether or not to use significance tests as part of scientific enquiry.~\cite{mayo-2022} argue that significance tests should not be abandoned. The ASA's statement on $p$-values is not against their use in scientific inquiry. In fact, Principle 1 states that ``a $p$-value can indicate how incompatible the data are with a specified statistical model''~\citep{wasserstein-2016}. The ASA statement calls for not relying solely on $p$-values to come to conclusions and make policy decisions. It also liberates investigators from one-size-fits-all thresholds. In any investigation, appropriate thresholds will have to be established in advance. As we explore ways to evaluate if replication rates are consistent with prevailing significance thresholds, let us also identify incentives for the diverse versions of unethical practices, remove such incentives and mete out heavy penalties for ethics violations.~\cite{nosek-2012} provide some ideas that can be implemented to promote credibility in science. As much as possible, fetters should be removed so that investigators can act and think for themselves. It is our belief that innovation thrives best under freedom.
\bibliographystyle{jofbib}
\bibliography{refs-wars}
\end{document}